\RequirePackage{ifpdf}

\documentclass[prl,showpacs,reprint,nofootinbib]{revtex4-1}

\usepackage[T1]{fontenc}
\usepackage{ae,aecompl}
\usepackage{bm}
\usepackage{latexsym}
\usepackage{dcolumn}
\usepackage{amsfonts,amssymb}
\usepackage{graphicx,epsfig}
\usepackage{psfrag}
\usepackage{amsmath,amssymb,multirow,array,mathtools}
\usepackage{ytableau}
\usepackage{tabularx}
\usepackage{float}
\usepackage{wrapfig}
\usepackage{caption}
\usepackage[parfill]{parskip}
\usepackage{subcaption}

\usepackage{hyperref} 
\hypersetup{
	colorlinks=true,	
	linkcolor=blue,		
	citecolor=red,		
	filecolor=blue,		
	urlcolor=blue		
}

\usepackage{cleveref}
\usepackage{soul, todonotes}

\newcommand{\Tr}{\text{Tr}}

\newcommand{\ben}{\begin{eqnarray}\displaystyle}
\newcommand{\een}{\end{eqnarray}}

\newcommand{\be}{\begin{equation}}
\newcommand{\ee}{\end{equation}}

\newcommand{\bc}{\begin{center}}
\newcommand{\ec}{\end{center}}

\newcommand{\eesp}{\end{split}}
\newcommand{\bsp}{\begin{split}}

\newcommand{\Rmnum}[1]{\expandafter\@slowromancap\romannumeral #1@}

\renewcommand{\l}{\lambda}	
\newcommand{\q}{\theta}	
	
\renewcommand{\r}{\rho}		

\newcommand{\cC}{\mathcal{C}}

\newcommand{\cK}{\mathcal{K}}

\newcommand{\cR}{\mathcal{R}}
\newcommand{\cS}{\mathcal{S}}
\newcommand{\cT}{\mathcal{T}}

\newcommand{\cV}{\mathcal{V}}

\newcommand{\cZ}{\mathcal{Z}}

\newcommand{\ra}{\rightarrow}

\newcommand{\lB}{\left [}
\newcommand{\rB}{\right ]}
\newcommand{\lb}{\left (}
\newcommand{\rb}{\right )}

\newcommand{\where}{\text{where}}

\newcommand{\bensp}{\begin{eqnarray}\begin{split}}
\newcommand{\eensp}{\end{eqnarray}\end{split}}

\newcommand{\bnm}{\begin{matrix}}
\newcommand{\enm}{\end{matrix}}

\def\Xint#1{\mathchoice
{\XXint\displaystyle\textstyle{#1}}%
{\XXint\textstyle\scriptstyle{#1}}%
{\XXint\scriptstyle\scriptscriptstyle{#1}}%
{\XXint\scriptscriptstyle\scriptscriptstyle{#1}}%
\!\int}
\def\XXint#1#2#3{{\setbox0=\hbox{$#1{#2#3}{\int}$ }
\vcenter{\hbox{$#2#3$ }}\kern-.6\wd0}}

\def\dashint{\Xint-}

\newcommand{\tl}{\tilde \lambda }

\begin{document}

\title{A New Phase in Chern-Simons Theory on Lens Space}
\author{Kushal Chakraborty} 
\email[]{kushal16@iiserb.ac.in}
\affiliation{Indian Institute of Science Education \&\ Research Bhopal, Bhopal Bypass, Bhopal 462066, India}
\author{Suvankar Dutta}
\email[]{suvankar@iiserb.ac.in}
\affiliation{Indian Institute of Science Education \&\ Research Bhopal, Bhopal Bypass, Bhopal 462066, India}

\begin{abstract}
We study $U(N)_k$ Chern-Simons theory on \emph{lens space} in Seifert framing and write down the partition function as a unitary matrix model. In the large $k$ and large $N$ limit the eigenvalue density satisfies an upper cap $\frac{1}{2\pi\lambda}$ where $\lambda=N/(k+N)=\text{fixed}$.  The eigenvalue density of the standard gapped phase saturates the upper cap at a critical value of $\lambda$ and cease to exist beyond that. We find a new phase (cap-gap phase) in this theory for $\lambda$ beyond the critical value and see that the on-shell free energy for the cap-gap phase is less than that of the gapped phase.
\end{abstract}

\maketitle


\emph{Introduction :} Being a topological theory the partition function (PF) of Chern-Simons (CS) theory in \emph{three} dimensions is a topological invariant. At the quantum level topological invariance is also preserved but at the expense of a choice of \emph{framing} \cite{Wittenjones,atiyah1990}. The PF of CS theory with gauge group $G$, rank $k$ on Seifert manifold
$M_{(g,p)}$ can be obtained by surgery from the expectation value of Wilson loop in $S^2\times S^1$. Different choices of surgery gives different framings in $M_{(g,p)}$. The PF on $M_{(g,p)}$ is given by \cite{Wittenjones}
\be\label{eq:cspfF}
\cZ(M_{(g,p)},G,k) = \sum_{\cR} \cK^{(p)}_{0\cR} \sum_{\cV} \cS_{0,\cV}^{1-2g}\cS_{\cR\cV}.
\ee
Here $\cK^{(p)}$ is a surgery/framing dependent matrix, $\cS_{\cR\cR'}$ is modular transform matrix associated with highest weight representations of affine Lie algebra $g_k$ of $G_k$ under inversion of modular parameter. The sum in (\ref{eq:cspfF}) runs over integrable representations of $g_k$ and $\cR =0$ corresponds to trivial representation. For $\cK^{(p)}=\cS \cT^{-p}\cS$ (Seifert framing), where $\cT_{\cR\cR'}$ is the second modular transform matrix associated with translation of modular parameter, the PF is given by
\ben\label{eq:cspfSF}
\cZ^{\text{SF}}(M_{(g,p)},G,k) = \sum_{\cR} \cS_{0,\cR}^{2-2g}\cT_{\cR\cR}^{-p}.
\een
Blau and Thompson \cite{Blau:2006gh} obtained the above PF using the method of \emph{abelianisation} \cite{blauthompson}. It turns out that their calculations renders the PF in Seifert framing. Using non-abelian localisation method one also obtains the PF in the same framing \cite{Beasley:2005vf}. We are interested in CS theory on lens space which is a Seifert manifold with genus $g=0$ and the first Chern class $p$. For $p=1$ the Seifert manifold is $S^3$. On $S^3$, there exists a canonical framing $\cK^{(p)}=S$ \cite{Wittenjones,Blau:2006gh} in which the PF is given by $\cZ^{\text{Can}}(S^3,G,k)=\cS_{00}$. Using the properties of $\cS$ and $\cT$ matrices ($\cS^2=(\cS \cT)^3=\mathbb{I}$) one can show that PF on $S^3$ in canonical and Seifert framings are related by :  $\cZ^{\text{SF}}(S^3,G,k)=\cT_{00}^2\cS_{00}$. In this paper we focus on the affine gauge group $G=U(N)_k$. In the large $k$ and large $N$ limit keeping
\be
\lambda=\frac{N}{k+N}
\ee
fixed, one can compute the PF in Seifert framing (\ref{eq:cspfSF}) under the saddle point approximation and find the dominant representation for $0\leq \l\leq 1$ \cite{Marino:2004eq, Marino:2002fk, Arsiwalla:2005jb,Aganagic:2002wv}. But the dominant representations fail to be integrable for all values of $\lambda$. It was first pointed out in \cite{Chattopadhyay:2019lpr}. In this paper we address this issue in detail.  Using the fact that the sum in (\ref{eq:cspfSF}) runs over integrable representations, we write the PF (for $g=0$) as a unitary matrix model. In the large $k,N$ limit the eigenvalue density is constrained to have a maximum value $\frac{1}{2\pi \lambda}$. We derive the saddle point equation for the eigenvalue density and find that the saddle point equation admits a gapped solution for $\lambda>0$. As a consistency check, we compute the PF on $S^3$ on the gapped solution and see that it is equal to $\cT_{00}^2\cS_{00}$ for all values of $0\leq\l\leq1$. Therefore the gapped phase is equivalent to the dominant representation obtained in \cite{Marino:2004eq, Marino:2002fk, Arsiwalla:2005jb,Aganagic:2002wv}. However, the eigenvalue density of the gapped phase saturates the upper bound at $\lambda=1/\pi \log\cosh\pi/p \equiv \lambda^*$ and seizes to exists beyond $\lambda^*$ \cite{Chattopadhyay:2019lpr}. In this paper we discover that there exists another phase (we call this phase cap-gap phase) for $\lambda>\lambda^*$ for $p\geq 1$. We compute the free energy of the cap-gap phase and see that it is less than that of the gapped phase for $\l>\l^*$. Therefore our calculation shows that the PF of CS theory on lens space (\ref{eq:cspfSF}) admits a phase transition at $\lambda=\lambda^*$ in the large $k,N$ limit when we consider the integrability condition properly. The advantage of converting the PF to a unitary matrix model is that the dominant representations for both the gapped and cap-gap phases are integrable by construction.

CS on lens space also enjoys the level-rank duality \cite{Naculich:2007nc}. We find the Young diagram (YD) distribution for a large $k,N$ phase and its dual and show that they are related by transposition followed by a shift. We also check that the PFs of dual theories are the same in the large $k,N$ limit. The dual of a gapped phase has an upper cap in the eigenvalue distribution. On the other hand, the cap-gap phase is dual to itself. This is similar to the matter CS theories on $S^2\times S^1$ studied by \cite{shirazs2s1,Chattopadhyay:2018wkp}.

\emph{$U(N)$ Chern-Simons theory on Seifert manifold :} The affine Lie algebra $u(N)_k$ is the quotient of $su(N)_k \times u(1)_{N(k+N)}$ by $\mathbb{Z}_N$. Hence $u(N)$ representation can be written in terms of $su(N)$ representations and eigenvalues of $u(1)$ generator : $\cR = (R,Q)$. We use the notation $R$ for $su(N)$ representations and $Q$ is eigenvalue of $u(1)$ generator, given by $Q= r(R)\ \text{mod} \ N$, where $r(R)$ is the number of boxes in $R$. Trivial representation $\cR=0$ corresponds to $R=0$ and $Q=0$. The modular transform matrix $\cS_{\cR\cR'}$ for $u(N)_k$ can be written in terms of representations of $su(N)$ and the $u(1)$ charges \cite{Naculich:2007nc,Naculich:1991,yellowbook}
\begin{eqnarray}\label{eq:Smod}
\cS_{\cR\cR'}=\frac{(-i)^{\frac{N(N-1)}2}}{(k+N)^{\frac N2}} e^{-\frac{2\pi i Q Q'}{ N(N+k)}}\det M(R,R')
\end{eqnarray}
where, $M(R,R')$ is a $N\times N$ matrix with elements,
\ben
M_{ij}(R,R') = \exp\lB\frac{2\pi i}{ k+N}\phi_i(R)\phi_j(R')\rB,
\een
\ben\label{eq:phii}
	\phi_i(R) = l_i-\frac{r(R)}{ N}-i-\frac{1}{2} (N+1)
\end{eqnarray}
and $l_i$'s are the number of boxes in $i^{th}$ row in $R$. The other modular transformation  matrix $\cT_{\cR \cR'}$ is given by
\ben
\cT_{\cR \cR'} & =e^{2\pi i(h_{R}-\frac{c}{24})} \delta_{\cR\cR'},\
h_{R} =\frac{1}{2}\frac{C_{2}(\cR)}{k+N},\
c =\frac{N(Nk+1)}{k+N}\ \ \
\een
where $\cC_2(\cR)$ is the quadratic Casimir of $u(N)_k$. Since $Q=r(R)+N s$ for $s\in \mathbb{Z}$, $u(N)$ representations $\cR$ can be characterised by extended YDs by re-defining number of boxes in $i^{th}$ row $\bar l_i=l_i+s$ for $1\leq i\leq N-1$ and $\bar l_N=s$. Now $\bar l_i$s can be negative and the corresponding YDs will have \emph{anti-boxes} \cite{Aganagic:2005dh}. In terms of these extended YDs the quadratic Casimir $C_2(\cR)$ is given by
\ben
C_2(\cR) = \sum_{i=1}^N \bar l_i(\bar l_i-2i+N+1).
\een
\emph{A representation $\cR$ of $u(N)_k$ is an integrable representation if $0\leq \bar l_N\leq \cdots \leq \bar l_1 \leq k$}  \cite{Naculich:2007nc}.

\emph{Chern-Simons theory as unitary matrix model :} For an integrable representation $\cR$ the hook numbers $h_i=\bar l_i +N-i$ satisfy $0< h_N < \cdots < h_1 \leq k+N$. Introducing new variables 
\begin{equation}\label{eq:thetaidef}
    \theta_{i} = \frac{2\pi}{N+K}\lb h_{i}-\frac{N-1}{2}\rb 
\end{equation}
we write the CS partition function (\ref{eq:cspfSF}) for $g=0$ in terms of $\theta_i$s ($\cZ^{\text{SF}}(M_{(0,p)},U(N),k) \ra \cZ_{N,k}^{p}$)
\ben\label{eq:pfS3}
\cZ_{N,k}^{p} & =&
\frac{1}{(N+k)^{N}}\sum_{\{\theta_{i}\}}\exp \Bigg[ \frac{1}{2}\sum_{i\neq j}^{N}\log\lB 4\sin^{2} \left( \frac{ \theta_{i}-\theta_{j}}{2} \right)\rB \nonumber \\  && - i p \left(\frac{N+k}{\pi} \sum_{i=1}^{N} \left( \frac{\theta_{i}^{2}}{4} - \frac{\pi^{2}}{12}\right) + \frac{\pi NK}{12}\right)\Bigg].
\een
The effective action (argument of the exponential) is symmetric in $\theta_i\ra -\theta_i$. Since the distribution of $\theta_i$s has a maximum range $2\pi$ (from eqn.(\ref{eq:thetaidef})) any classical configuration will satisfy $-\pi \leq \theta_i \leq \pi$. The potential is neither real nor periodic. However, in order to write a unitary matrix model for CS theory we demand that $\theta_i$s are periodic with periodicity $2\pi$ which essentially means that we impose periodicity in hook numbers $\bar h_i$s : $\bar h_i\sim \bar h_i + k + N$. To make the potential real we analytically continue $p\ra - i p$. This allows us to write the above partition function as a unitary matrix model with potential $\sim \sum_{n>0}\frac{(-1)^2}{2n^2}\lb \Tr U^n +\Tr U^{\dagger n}\rb$ \cite{Okuda}. Later we see that in the large $k,N$ limit the on-shell partition function on $S^3$ matches with $\cS_{00}$ (up to a phase) when we carefully replace $p\ra i p$.

In the continuum limit we define an eigenvalue density $\rho(\theta)=\frac{1}{N}\sum_{i=1}^{N}\delta(\theta-\theta_{i})$. The partition function is given by
\ben\label{eq:se}
\begin{split}
\cZ_{N,k}^{p}   & = \int [d\theta]e^{-(N+K)^{2}S_{eff}[\rho]}, \quad \where  \\
        S_{eff}[\rho] & = \frac{p \lambda}{\pi}\int \rho(\theta)\lb \frac{\theta^{2}}{4} - \frac{\pi^{2}}{12} \rb d \theta + \frac{\pi p \lambda(1-\lambda)}{12} \\ 
        & - \frac{\lambda^{2}}{2}\int \dashint\rho(\theta)\rho(\theta')\log\lB 4\sin^{2}\lb \frac{\theta-\theta'}{2}\rb \rB d\theta d\theta'
        \end{split}.
\een
The saddle point equation for $\rho(\theta)$, obtained from this effective action is given by
\begin{equation}\label{eq:sad}
          \dashint\rho(\theta')\cot\left(\frac{\theta-\theta'}{2}\right)d\theta'=\frac{p}{2\pi\lambda}\theta .
\end{equation}
From the definition of $\theta_i$s (\ref{eq:thetaidef}) we see that the minimum separation between $\theta_i$ and $\theta_{i+1}$ is $2\pi/(N+k)$. This implies that in the large $k,N$ limit the eigenvalue density $\rho(\theta)$ satisfies an upper bound
\ben\label{eq:rhobound}
\rho(\theta)\leq \frac1{2\pi \lambda}.
\een
Therefore we have to solve the saddle point equation (\ref{eq:sad}) for $\rho(\theta)$ in presence of this constraint. Note that the YD distribution can be obtained from the eigenvalue distribution : $u(h)=2\pi \lambda \rho(\theta)$. Hence, $\rho(\theta)$ having an upper cap $1/2\pi\lambda$ means $u(h)\leq 1$. Before we discuss the large $k,N$ phases of this theory we take a pause to study the eigenvalue density of the level-rank dual theory and its connection to $\rho(\theta)$.

\emph{Level-Rank duality :} $N\leftrightarrow k$ duality in $U(N)$ CS theory implies that the dominant YDs in two theories, dual to each other are related by a transposition followed by a shift. In order to prove this statement we write down the partition function of $U(N)_k$ CS theory in terms of number of boxes in different columns in a YD. A YD corresponding to an integrable representation of $u(N)_k$ can be characterised by $\bar v_\mu$ - the number of boxes in $\mu^{th}$ column of a YD $\cR$ where $1\leq \mu \leq k$ and $\bar v_1\leq N$. $\{\bar v_\mu\}$ is the set of box numbers in different rows of $\tilde \cR$, where $\tilde \cR$ is transpose of $\cR$. The quadratic Casimir $C_2(\cR)$ can be written in terms of $\bar v_\mu$. Also the $\cS$ modular transform matrix (\ref{eq:Smod}) is invariant under transposition \cite{Naculich:2007nc}. We introduce new variables 
\ben\label{eq:phimudef}
\phi_\mu  = \frac{2\pi}{k+N}\lb w_\mu -\frac{k+N-1}{2}\rb, \ w_\mu = \bar v_\mu +k -\mu \hspace{.6cm}
\een
Since $0\leq \bar v_\mu \leq N$, $\phi_\mu$s are distributed in a range of $2\pi$. The partition function (\ref{eq:cspfSF}) can be written in terms of $\phi_\mu$s and it turns out that the effective action is symmetric under $\phi_\mu \ra 2\pi -\phi_\mu$. Hence for any classical solution $\phi_\mu$s are distributed symmetrically about $\phi=\pi$ from $0$ to $2\pi$. In the continuum limit we define a distribution functions for $\phi_\mu$s
\ben
\tilde{\rho}(\phi)=\frac{1}{k}\sum_{\mu=1}^{k}\delta(\phi-\phi_{\mu})
\een
and the partition function is given by
\begin{equation}
    \cZ_{N,k}^p=\int [d\phi]e^{-(N+k)^{2}\tilde{S}_{eff}[\tilde{\rho}]}
\end{equation}
where,
\ben\label{eq:sed}
        && \tilde{S}_{eff}[\tilde{\rho}] =  \frac{p \tl }{\pi}\int \tilde{\rho}(\phi) \lb \frac{\pi^{2}}{12} - \frac{(\phi-\pi)^{2}}{4} \rb d\phi + \frac{p \pi \lambda \tl }{12} \nonumber \\ && \quad \ \ - \frac{\tl^{2}}{2}\int\dashint \tilde{\rho}(\phi) \tilde{\rho}(\phi') \log\lb 4\sin^{2}(\frac{\phi-\phi'}{2})\rb d\phi d\phi'
\een
and $\tl=1-\lambda$. The saddle point equation for $\tilde\rho(\phi)$ is given by
\ben\label{eq:sadrhotilde}
\dashint_{0}^{2\pi}\tilde{\rho}(\phi')\cot\lb \frac{\phi-\phi'}{2} \rb d\phi' = \frac{p}{2\pi\tl}\lb \pi - \phi \rb.
\een
Comparing (\ref{eq:sad}) and (\ref{eq:sadrhotilde}) we find
\begin{equation}\label{eq:dend}
    \tilde{\rho}(\phi)=\frac{1}{2\pi\tl}-\frac{\lambda }{\tl}\rho(\phi+\pi).
\end{equation}
This relation establishes the fact that under $N\leftrightarrow k$ duality the dominant YDs in $U(N)_k$ and $U(k)_N$ CS theories are related by a transposition with a shift $\frac{N+k}{2}$. Using (\ref{eq:dend}) we also see that $S[\rho]=\tilde S[\tilde \rho]$. The relation (\ref{eq:dend}) is similar to what found in \cite{shirazs2s1,Chattopadhyay:2019lpr} in the context of matter CS theory on $S^2\times S^1$.

\begin{figure}[h]
\centering
\includegraphics[width=6cm,height=4cm]{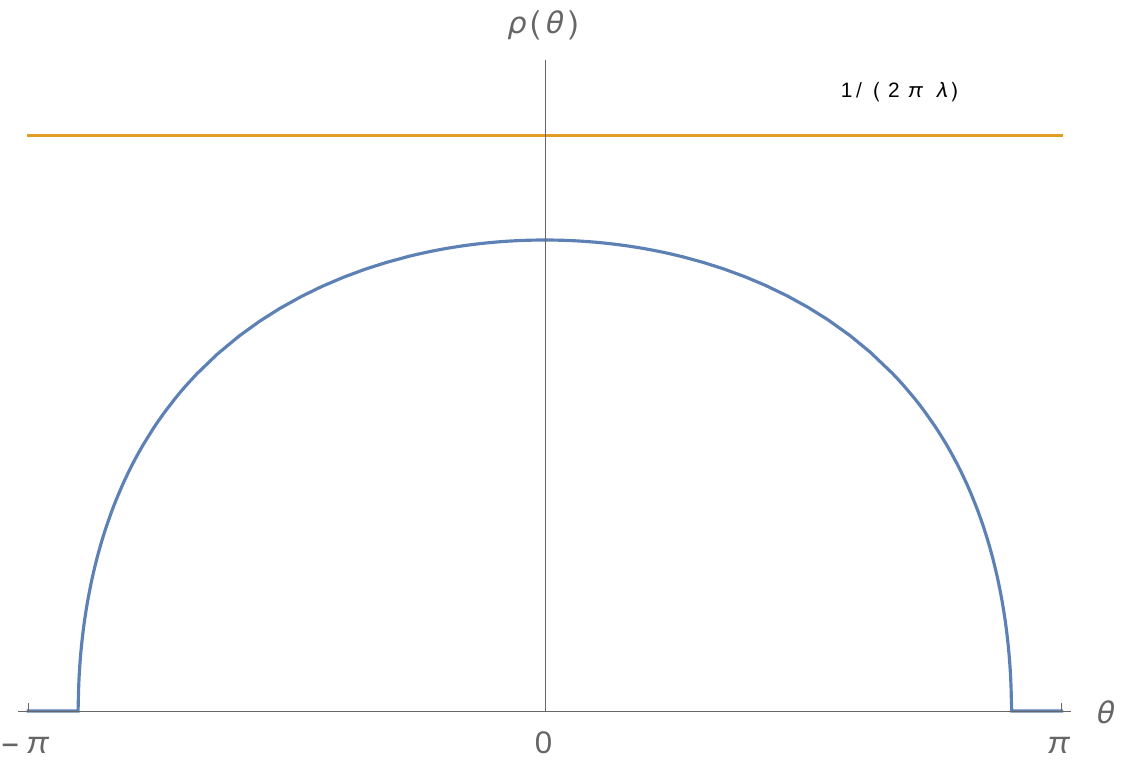}
\caption{$\rho(\theta)$ for one-gap phase.}
\label{fig:onegap}
\end{figure}

\emph{Large $N$ phases :} The unitary matrix model (\ref{eq:se}) was studied in \cite{Chattopadhyay,Chattopadhyay:2019lpr}. It was observed that the system has a gapped phase in the large $k,N$ limit and the eigenvalue distribution is given by,
\begin{eqnarray}\label{eq:csev1gap}
\r(\theta) =
\frac{p}{2\pi^2 \l}\tanh^{-1}\left[ 
\sqrt{1- \frac{e^{-\frac{2\pi\l}{p}}}{\cos^{2}\frac{\theta}{2}}}\right].
\end{eqnarray}
Since $\r(\q)\geq 0$, this implies eigenvalues are distributed over the range
\begin{equation}
-2\cos^{-1}e^{-\frac{\pi \l}{p}}<\theta< 2\cos^{-1}e^{-\frac{\pi\l}{p}}.
\end{equation}
See fig.\ref{fig:onegap} for the eigenvalue distribution.
We calculate the PF (\ref{eq:pfS3}) on $S^3$ (i.e. for $p=1$) on this solution (\ref{eq:csev1gap}) and check that after suitable analytic continuation $p\ra i p$ (before setting $p=1$) the PF exactly matches with $\cT_{00}^{2}\cS_{00}$ for all values of $0\leq \l\leq 1$. Hence this phase is equivalent to the dominant phase obtained in \cite{Marino:2004eq, Marino:2002fk, Arsiwalla:2005jb,Aganagic:2002wv}. However, due to the constraint (\ref{eq:rhobound}) on $\rho(\theta)$ the eigenvalue density saturates the upper bound at $\lambda= 1/\pi \log\cosh(\pi/p) \equiv \lambda^* $ \cite{Chattopadhyay:2019lpr}. Therefore the gapped phase is not valid anymore for $\lambda>\lambda^*$ for any $p\geq 1$.

\begin{figure}[h]
\centering
\includegraphics[width=6cm,height=4cm]{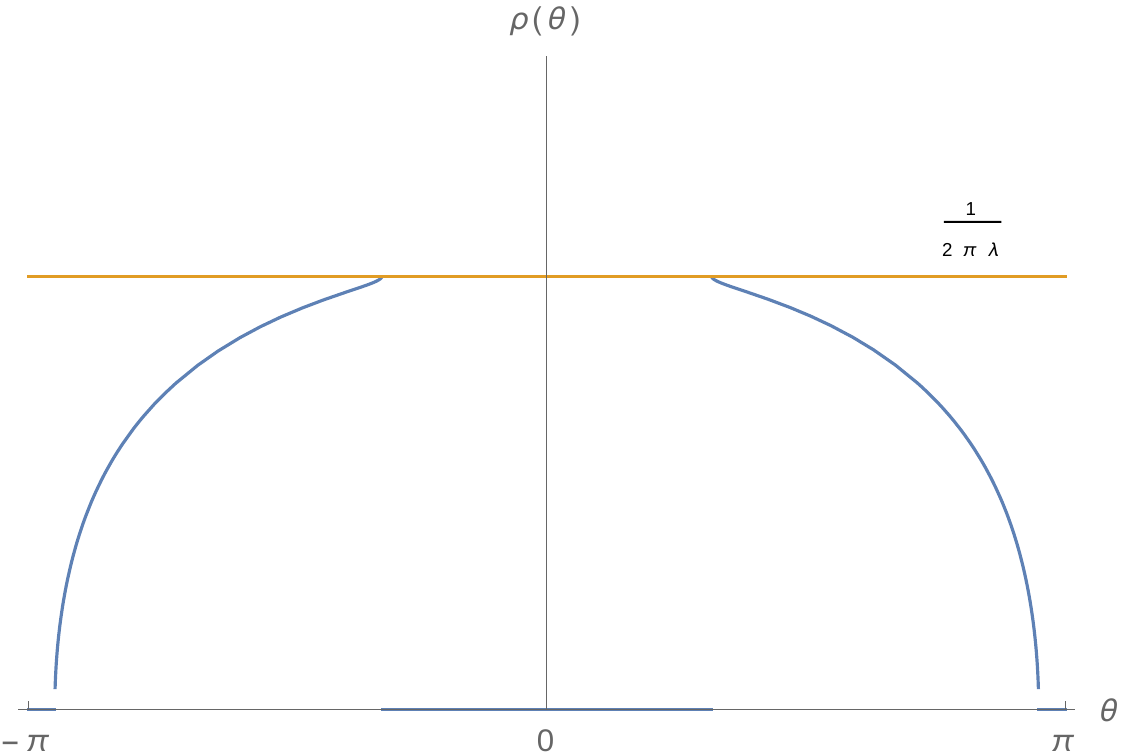}
\caption{$\rho(\theta)$ for cap-gap phase.}
\label{fig:capgap}
\end{figure}

\begin{figure}[h]
\centering
\includegraphics[width=6cm,height=4cm]{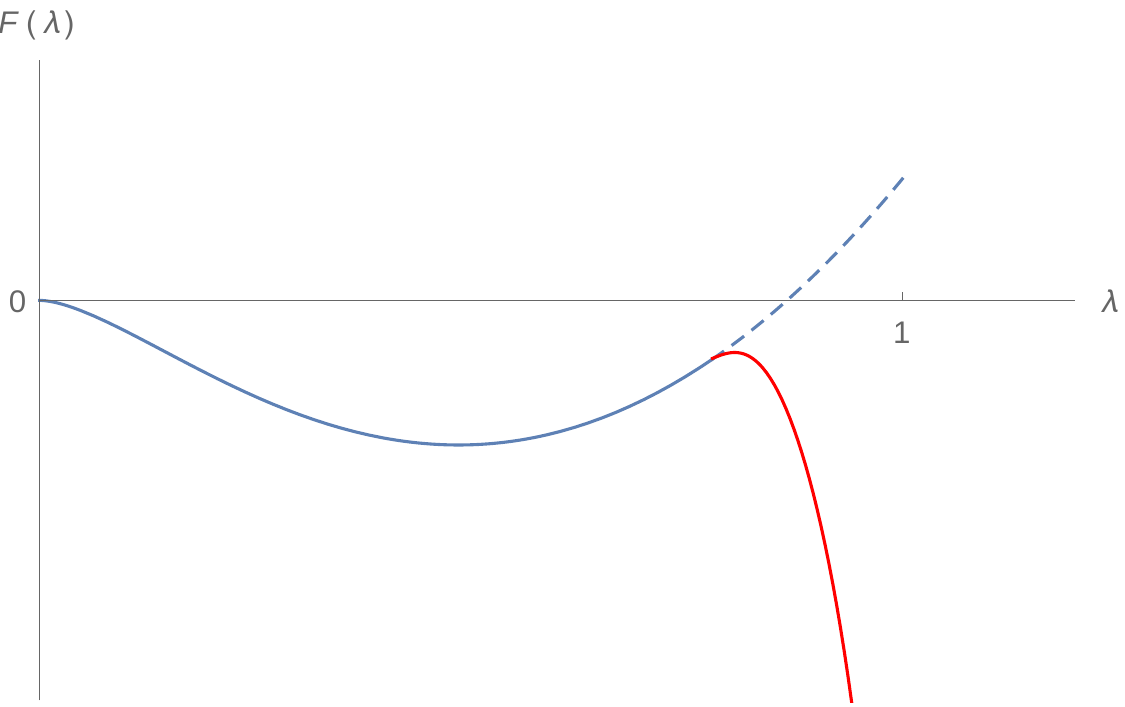}
\caption{Free energy of CS on $S^3$ as a function of $\lambda$. The solid blue line is the free energy for the gapped phase for $\l<\l^*$. The dashed blue line is the extension of the same beyond $\lambda^*$. The red line depicts the free energy for the cap-gap phase.}
\label{fig:free}
\end{figure}

\emph{Cap-gap phase :} For $\lambda>\lambda^*$ the eigenvalue density develops a \emph{cap} about $\theta=0$. To find that phase we take the following ansatz for $\rho(\theta)$
 \ben
   \rho(\theta) = \Bigg\{ {\frac{1}{2\pi \lambda}  \hspace{.3cm} \text{for}\  -\theta_{2}<\theta<\theta_{2} \hfill \atop
     \hat\rho(\theta)\ \ \text{for}\  -\theta_{1}<\theta<-\theta_{2}\  \text{and}\  \theta_{2}<\theta<\theta_{1}. }
 \een
Using the map $z=e^{i\theta}$, the saddle point equation for $\hat \rho(\theta)$ is given by
\begin{equation}
\dashint\hat\rho(z')\frac{z+z'}{z-z'}dz'=\frac{p\log(z)}{2\pi i\lambda}-\frac{1}{2\pi\lambda}\int\frac{1}{z'}\frac{z+z'}{z-z'}dz'.
\end{equation}
Following \cite{shirazs2s1} we define a resolvent function $\Phi(z)$
\ben\label{eq:Phiz}
         \Phi(z)&=&\int\frac{\hat\rho(z')}{i z'}\frac{z+z'}{z-z'}dz'=h(z)H(z), \quad \where \\
h(z)&=&\sqrt{(z^{2}-2z\cos \theta_{1}+1)(z^{2}-2z \cos \theta_{2}+1)}. \nonumber
\een
From the normalization of eigenvalue density it follows that,
\ben\label{eq:Phiasym}
\Phi(z\to\infty) & \sim  & 1 - \frac{1}{2\pi\lambda} \int \frac{d\omega}{i\omega}.
\een
The resolvent $\Phi(z)$ has branch cut in complex $z$ plane. The eigenvalue density $\hat \rho(z)$ is obtained from the discontinuity of $\Phi(z)$
\be
\Phi^{+}(z)-\Phi^{-}(z) = 4\pi \hat{\rho}(z).
\ee
Following \cite{Migdal:1984gj} the function $H(z)$ can be evaluated as
\ben\label{eq:Hz}
H(z)=i \oint \frac{dw}{2\pi i} \frac{\frac{p\log(w)}{2\pi i\lambda} - \frac{1}{2\pi\lambda}\int\frac{1}{s} \frac{w+s}{w-s}ds}{h(w)(w-z)}.
\een
Plugging this expression in (\ref{eq:Phiz}) we expand the r.h.s. for large $z$ and comparing the expression with (\ref{eq:Phiasym}) we find the following two constraint
 \ben\label{eqab}
         \frac{p}{2\pi\lambda}\oint\frac{dz}{2\pi i}\frac{\log(z)}{h(z)}+\frac{i}{\pi\lambda}\int\frac{d\omega}{h(\omega)} & =& 0,\nonumber\\
        1 + \frac{p}{2\pi\lambda}\oint\frac{dz}{2\pi i}\frac{z\log(z)}{h(z)}+\frac{i}{\pi\lambda}\int\frac{\omega d\omega}{h(\omega)} & =& 0.
 \een
Using the formula given in appendix, we numerically solve these two equations to find the endpoints $\theta_1$ and $\theta_2$. From the discontinuity $\Phi(z)$ we compute the eigenvalue density
\ben
\hat{\rho}(\theta) &=& - \frac{|\sin\phi|}{\pi^{2}\lambda}\frac{\sqrt{(\sin^{2}\frac{\phi}{2}-\sin^{2}\frac{\theta_{2}}{2})(\sin^{2}\frac{\theta_{1}}{2}-\sin^{2}\frac{\phi}{2})}}{\sqrt{(1+\cos\theta_{2})(1-\cos\theta_{1})}}\nonumber \\
&& \quad \Bigg[\frac{2p \lb \cos^{2}\frac{\theta_{1}}{2} \Pi(\psi,n_{1},m_{1}) - \cos^{2}\frac{\phi}{2} F(\psi,m_{1})\rb} {(1+\cos\phi) (\cos\phi-\cos\theta_{1})}\nonumber\\   && \quad -\frac{4\lb \Pi(n_{2},m_{2})-\sin^{2}\frac{\phi}{2}K(m_{2})\rb}{\sin^{2}\phi}\Bigg]
\een
where $m_1,m_2,n_1,n_2, \psi$ are given in (\ref{eq:appn1n2}). The eigenvalue density in cap-gap phase is plotted in fig. \ref{fig:capgap}.
The free energies for these two phases (for $p=1$) as a function of $\lambda$ are plotted in fig. \ref{fig:free}. From this figure we see that the cap-gap phase has free energy less than that of gapped phase and hence dominant over the gapped phase for $\lambda>\lambda^*$. 
However the on-shell $S^3$ PF on the cap-gap phase differs from $\cT_{00}^2\cS_{00}$.

\emph{Discussion :} In order to obtain a real saddle point equation (\ref{eq:sad}) we use an analytic continuation in $p$. However if we use an analytic continuation in $\lambda$ \cite{Chattopadhyay:2019lpr}, we would also get a real saddle point equation for YD distribution $u(h)$ in $h$ plane with a $\coth$ kernel similar to what considered in \cite{Arsiwalla:2005jb}. Solution of this equation renders a YD distribution which crosses the maximum value $1$ for $p>2$ only. Hence one can exhibit a phase transition only for $p\geq 3$. The YD distribution is different from the one-gap eigenvalue distribution obtained in this paper. But with a proper analytic continuation of $p$ and $\lambda$ one can relate the two \cite{Chattopadhyay:2019lpr}. However, the YD distribution obtained in \cite{Arsiwalla:2005jb,Marino:2004eq,Chattopadhyay:2019lpr} violates the integrability bound for some value of $\lambda$ between $0$ and $1$. In strictly $k\ra \infty$ limit the sum over $\cR$ in (\ref{eq:cspfSF}) is unrestricted. Hence one should not expect any phase transition in the system. In this paper we have shown that calculating free energy in saddle point approximation matches with $T_{00}^2 S_{00}$ after we analytically continue $p$ back to its original value.
One can look into the analytic structure of free energy in the complex $p$ plane and see that there is absolutely no problem in going from $p\leftrightarrow ip$. But the same trick does not give the same result in the other phase of the theory. The question is why. To understand this we need to look at the relation (\ref{eq:cspfSF}). The sum is over integrable representations. This is valid for any $N$ and $k$ (however large). If one takes $N, k \ra \infty $ limit without any restriction, then the sum runs over all possible Young diagrams with any number of rows and any number of columns. However, here we are considering a particular limit $N, k \ra \infty $ keeping $N/k$ fixed. Under this condition the sum becomes restricted - one does not sum over all possible Young diagrams. Therefore we do not expect that in the double scaling limit the above identity holds. This is exactly what we see in our derivation. In the double scaling limit  $\theta_i$s are defined in such a way (\ref{eq:thetaidef}) that they have range between $0$ and $2\pi$ and the dominant representations are always integrable. But this change of variables imposes a cap on the eigenvalue distribution which triggers a phase transition in the theory.

The 't Hooft expansion of the PF of $SU(N)$ CS theory on $S^3$ is proposed to be dual to topological closed string theory on the $S^2$ blow up of the conifold geometry \cite{Gopakumar:1998ki} for arbitrary $\lambda$ and all orders of $1/N$. In canonical framing the CS PF is equal to $\cS_{00}$ and an exact function of $\lambda$ which matches with string theory side. In Seifert framing, we observe that the PF of CS theory in the gapped phase is equal to that in the string theory side. But the PF in cap-gap phase differs from $\cS_{00}$ for $\lambda>\lambda^*$. Dependence of phase on the choice of framing is bit puzzling here. The question is why a new phase pops up in the theory when we take the double scaling limit. The saddle equation (\ref{eq:sad}) also admits multi-cut solutions, which were related to some non-perturbative D-instantons \cite{Morita:2017oev}, are different than the cap-gap phase. It would be interesting to understand the meaning of this new phase in the string theory side as well.

We explicitly check the level-rank duality in CS theory on $S^3$. The theory admits three types of phases. For $\lambda<\lambda^*$ one has gapped phase and capped phase. These two phases are level-rank dual to each other. For $\lambda>\lambda^*$ the theory admits a cap-gap phase which is level rank dual to itself. There is a third order phase transition at $\lambda^*$. The phase structure is similar to that of CS-matter theory on $S^2\times S^1$ \cite{shirazs2s1,Chattopadhyay:2019lpr} except that here we do not have any gap less phase. 

The partition function of $q$-deformed $U(N)$ Yang-Mills on a generic Riemann surface with \emph{zero} $\theta$ term is equal to the PF of CS theory on $M_{(g,p)}$ up to a phase factor for $q=e^{\frac{2\pi i}{N+k}}$ and $k,p \in \mathbb{Z}$ \cite{Naculich:2007nc}. Thus our analysis shows that the $q$-deformed Yang-Mills undergoes a phase transition even for $p=1$ unlike \cite{Arsiwalla:2005jb}.

{\bf Acknowledgments:} We thank Arghya Chattopadhyay and Neetu for working on this problem at the initial stage. We are grateful to Rajesh Gopakumar and Dileep Jatkar for reading our manuscript and giving their valuable comments. The work of SD is supported by the \emph{MATRICS} grant (no. \emph{MTR/2019/000390}, the Department of Science and Technology, Government of India). We are indebted to people of India for their unconditional support toward the researches in basic science.


\emph{Appendix - Useful formula : } We use the following useful results in our calculations.

\ben
&& \oint\frac{dz}{2\pi i}\frac{\log(z)}{h(z)} = \frac{2 F\lb \psi,m_{1}\rb }{\sqrt{(1+\cos\theta_{2}) (1-\cos\theta_{1})}}\nonumber \\
&& \oint\frac{dz}{2\pi i} \frac{z\log(z)}{h(z)} = \frac{2\cos\theta_{1}  F \lb \psi, m_{1}\rb }{\sqrt{(1+\cos\theta_{2}) (1-\cos\theta_{1})}} +  \frac{2 \beta  v_{4}'(\beta)}{v_{4}(\beta)}
          \nonumber \\
         && + \frac{1}{2}\log\lb \frac{(1-\cos\theta_{1})(1+\cos\theta_{2})}{4K^2 (m_{1})}\rb -\log\left[\frac{v_{1}(2\beta)}{v_{1}'(0)}\right]
 \een
where,
\ben
\beta = \frac{F(\psi,m_{1})}{2K(m_{1})} ,\
q = e^{-\pi\frac{K'(m_{1})}{K(m_{1})}},\ K'(m_1) = K(\sqrt{1-m_{1}^{2}}).\nonumber
\een
 \ben
     \int_{e^{-i\theta_{2}}}^{e^{i\theta_{2}}}\frac{d\omega}{h(\omega)} &=& \frac{2 i K\lb m_{2}\rb }{\sqrt{(1-\cos\theta_{1})(1+\cos\theta_{2})}}, \nonumber\\
      \int_{e^{-i\theta_{2}}}^{e^{i\theta_{2}}}\frac{\omega d\omega}{h(\omega)} &=& \frac{2 i\lb 2\Pi\lb n,m_{2}\rb-K\lb m_{2}\rb \rb }{\sqrt{(1-\cos\theta_{1})(1+\cos\theta_{2})}}
 \een
 \ben\label{eq:appn1n2}
   && \psi= \sin^{-1}\sqrt{\frac{1-\cos\theta_{1}}{2}},\  m_{1}= \sqrt{ \frac{2(\cos\theta_{2}-\cos\theta_{1})}{(1+\cos\theta_{2})(1-\cos\theta_{1})}}, \nonumber \\ &&
    m_{2}=\sqrt{\frac{(1-\cos\theta_{2})(1+\cos\theta_{1})}{(1-\cos\theta_{1})(1+\cos\theta_{2})}}; \ n=\frac{\cos\theta_{2}-1}{1+\cos\theta_{2}}; \\ && 
    n_{1}=\frac{2(\cos\phi-\cos\theta_{1})}{(1-\cos\theta_{1})(1+\cos\phi)};\ n_{2}=\frac{(1-\cos\theta_{2})(1+\cos\phi)}{(1+\cos\theta_{2})(1-\cos\phi)}\nonumber
  \een

\bibliography{Bib.bib}

\begin{thebibliography}{20}%
\makeatletter
\providecommand \@ifxundefined [1]{%
 \@ifx{#1\undefined}
}%
\providecommand \@ifnum [1]{%
 \ifnum #1\expandafter \@firstoftwo
 \else \expandafter \@secondoftwo
 \fi
}%
\providecommand \@ifx [1]{%
 \ifx #1\expandafter \@firstoftwo
 \else \expandafter \@secondoftwo
 \fi
}%
\providecommand \natexlab [1]{#1}%
\providecommand \enquote  [1]{``#1''}%
\providecommand \bibnamefont  [1]{#1}%
\providecommand \bibfnamefont [1]{#1}%
\providecommand \citenamefont [1]{#1}%
\providecommand \href@noop [0]{\@secondoftwo}%
\providecommand \href [0]{\begingroup \@sanitize@url \@href}%
\providecommand \@href[1]{\@@startlink{#1}\@@href}%
\providecommand \@@href[1]{\endgroup#1\@@endlink}%
\providecommand \@sanitize@url [0]{\catcode `\\12\catcode `\$12\catcode
  `\&12\catcode `\#12\catcode `\^12\catcode `\_12\catcode `\%12\relax}%
\providecommand \@@startlink[1]{}%
\providecommand \@@endlink[0]{}%
\providecommand \url  [0]{\begingroup\@sanitize@url \@url }%
\providecommand \@url [1]{\endgroup\@href {#1}{\urlprefix }}%
\providecommand \urlprefix  [0]{URL }%
\providecommand \Eprint [0]{\href }%
\providecommand \doibase [0]{http://dx.doi.org/}%
\providecommand \selectlanguage [0]{\@gobble}%
\providecommand \bibinfo  [0]{\@secondoftwo}%
\providecommand \bibfield  [0]{\@secondoftwo}%
\providecommand \translation [1]{[#1]}%
\providecommand \BibitemOpen [0]{}%
\providecommand \bibitemStop [0]{}%
\providecommand \bibitemNoStop [0]{.\EOS\space}%
\providecommand \EOS [0]{\spacefactor3000\relax}%
\providecommand \BibitemShut  [1]{\csname bibitem#1\endcsname}%
\let\auto@bib@innerbib\@empty
\bibitem [{\citenamefont {Witten}(1989)}]{Wittenjones}%
  \BibitemOpen
  \bibfield  {author} {\bibinfo {author} {\bibfnamefont {E.}~\bibnamefont
  {Witten}},\ }\href {\doibase 10.1007/BF01217730} {\bibfield  {journal}
  {\bibinfo  {journal} {Commun. Math. Phys.}\ }\textbf {\bibinfo {volume}
  {121}},\ \bibinfo {pages} {351} (\bibinfo {year} {1989})}\BibitemShut
  {NoStop}%
\bibitem [{\citenamefont {Atiyah}(1990)}]{atiyah1990}%
  \BibitemOpen
  \bibfield  {author} {\bibinfo {author} {\bibfnamefont {M.}~\bibnamefont
  {Atiyah}},\ }\href {\doibase https://doi.org/10.1016/0040-9383(90)90021-B}
  {\bibfield  {journal} {\bibinfo  {journal} {Topology}\ }\textbf {\bibinfo
  {volume} {29}},\ \bibinfo {pages} {1 } (\bibinfo {year} {1990})}\BibitemShut
  {NoStop}%
\bibitem [{\citenamefont {Blau}\ and\ \citenamefont
  {Thompson}(2006)}]{Blau:2006gh}%
  \BibitemOpen
  \bibfield  {author} {\bibinfo {author} {\bibfnamefont {M.}~\bibnamefont
  {Blau}}\ and\ \bibinfo {author} {\bibfnamefont {G.}~\bibnamefont
  {Thompson}},\ }\href {\doibase 10.1088/1126-6708/2006/05/003} {\bibfield
  {journal} {\bibinfo  {journal} {JHEP}\ }\textbf {\bibinfo {volume} {05}},\
  \bibinfo {pages} {003} (\bibinfo {year} {2006})},\ \Eprint
  {http://arxiv.org/abs/hep-th/0601068} {arXiv:hep-th/0601068 [hep-th]}
  \BibitemShut {NoStop}%
\bibitem [{\citenamefont {{Blau}}\ and\ \citenamefont
  {{Thompson}}(1993)}]{blauthompson}%
  \BibitemOpen
  \bibfield  {author} {\bibinfo {author} {\bibfnamefont {M.}~\bibnamefont
  {{Blau}}}\ and\ \bibinfo {author} {\bibfnamefont {G.}~\bibnamefont
  {{Thompson}}},\ }\href {\doibase 10.1016/0550-3213(93)90538-Z} {\bibfield
  {journal} {\bibinfo  {journal} {Nuclear Physics B}\ }\textbf {\bibinfo
  {volume} {408}},\ \bibinfo {pages} {345} (\bibinfo {year} {1993})},\ \Eprint
  {http://arxiv.org/abs/hep-th/9305010} {hep-th/9305010} \BibitemShut {NoStop}%
\bibitem [{\citenamefont {Marino}(2004{\natexlab{a}})}]{Marino:2004eq}%
  \BibitemOpen
  \bibfield  {author} {\bibinfo {author} {\bibfnamefont {M.}~\bibnamefont
  {Marino}}\ }(\bibinfo {year} {2004})\ \Eprint
  {http://arxiv.org/abs/hep-th/0410165} {arXiv:hep-th/0410165 [hep-th]}
  \BibitemShut {NoStop}%
\bibitem [{\citenamefont {Marino}(2004{\natexlab{b}})}]{Marino:2002fk}%
  \BibitemOpen
  \bibfield  {author} {\bibinfo {author} {\bibfnamefont {M.}~\bibnamefont
  {Marino}},\ }\href {\doibase 10.1007/s00220-004-1194-4} {\bibfield  {journal}
  {\bibinfo  {journal} {Commun. Math. Phys.}\ }\textbf {\bibinfo {volume}
  {253}},\ \bibinfo {pages} {25} (\bibinfo {year} {2004}{\natexlab{b}})},\
  \Eprint {http://arxiv.org/abs/hep-th/0207096} {arXiv:hep-th/0207096 [hep-th]}
  \BibitemShut {NoStop}%
\bibitem [{\citenamefont {Arsiwalla}\ \emph {et~al.}(2006)\citenamefont
  {Arsiwalla}, \citenamefont {Boels}, \citenamefont {Marino},\ and\
  \citenamefont {Sinkovics}}]{Arsiwalla:2005jb}%
  \BibitemOpen
  \bibfield  {author} {\bibinfo {author} {\bibfnamefont {X.}~\bibnamefont
  {Arsiwalla}}, \bibinfo {author} {\bibfnamefont {R.}~\bibnamefont {Boels}},
  \bibinfo {author} {\bibfnamefont {M.}~\bibnamefont {Marino}}, \ and\ \bibinfo
  {author} {\bibfnamefont {A.}~\bibnamefont {Sinkovics}},\ }\href {\doibase
  10.1103/PhysRevD.73.026005} {\bibfield  {journal} {\bibinfo  {journal} {Phys.
  Rev.}\ }\textbf {\bibinfo {volume} {D73}},\ \bibinfo {pages} {026005}
  (\bibinfo {year} {2006})},\ \Eprint {http://arxiv.org/abs/hep-th/0509002}
  {arXiv:hep-th/0509002 [hep-th]} \BibitemShut {NoStop}%
\bibitem [{\citenamefont {Aganagic}\ \emph {et~al.}(2004)\citenamefont
  {Aganagic}, \citenamefont {Klemm}, \citenamefont {Marino},\ and\
  \citenamefont {Vafa}}]{Aganagic:2002wv}%
  \BibitemOpen
  \bibfield  {author} {\bibinfo {author} {\bibfnamefont {M.}~\bibnamefont
  {Aganagic}}, \bibinfo {author} {\bibfnamefont {A.}~\bibnamefont {Klemm}},
  \bibinfo {author} {\bibfnamefont {M.}~\bibnamefont {Marino}}, \ and\ \bibinfo
  {author} {\bibfnamefont {C.}~\bibnamefont {Vafa}},\ }\href {\doibase
  10.1088/1126-6708/2004/02/010} {\bibfield  {journal} {\bibinfo  {journal}
  {JHEP}\ }\textbf {\bibinfo {volume} {02}},\ \bibinfo {pages} {010} (\bibinfo
  {year} {2004})},\ \Eprint {http://arxiv.org/abs/hep-th/0211098}
  {arXiv:hep-th/0211098 [hep-th]} \BibitemShut {NoStop}%
\bibitem [{\citenamefont {Chattopadhyay}\ \emph {et~al.}(2019)\citenamefont
  {Chattopadhyay}, \citenamefont {Suvankar},\ and\ \citenamefont
  {Neetu}}]{Chattopadhyay:2019lpr}%
  \BibitemOpen
  \bibfield  {author} {\bibinfo {author} {\bibfnamefont {A.}~\bibnamefont
  {Chattopadhyay}}, \bibinfo {author} {\bibfnamefont {D.}~\bibnamefont
  {Suvankar}}, \ and\ \bibinfo {author} {\bibnamefont {Neetu}},\ }\href
  {\doibase 10.1103/PhysRevD.100.126009} {\bibfield  {journal} {\bibinfo
  {journal} {Phys. Rev. D}\ }\textbf {\bibinfo {volume} {100}},\ \bibinfo
  {pages} {126009} (\bibinfo {year} {2019})},\ \Eprint
  {http://arxiv.org/abs/1902.07538} {arXiv:1902.07538 [hep-th]} \BibitemShut
  {NoStop}%
\bibitem [{\citenamefont {Naculich}\ and\ \citenamefont
  {Schnitzer}(2007)}]{Naculich:2007nc}%
  \BibitemOpen
  \bibfield  {author} {\bibinfo {author} {\bibfnamefont {S.~G.}\ \bibnamefont
  {Naculich}}\ and\ \bibinfo {author} {\bibfnamefont {H.~J.}\ \bibnamefont
  {Schnitzer}},\ }\href {\doibase 10.1088/1126-6708/2007/06/023} {\bibfield
  {journal} {\bibinfo  {journal} {JHEP}\ }\textbf {\bibinfo {volume} {06}},\
  \bibinfo {pages} {023} (\bibinfo {year} {2007})},\ \Eprint
  {http://arxiv.org/abs/hep-th/0703089} {arXiv:hep-th/0703089 [HEP-TH]}
  \BibitemShut {NoStop}%
\bibitem [{\citenamefont {Jain}\ \emph {et~al.}(2013)\citenamefont {Jain},
  \citenamefont {Minwalla}, \citenamefont {Sharma}, \citenamefont {Takimi},
  \citenamefont {Wadia},\ and\ \citenamefont {Yokoyama}}]{shirazs2s1}%
  \BibitemOpen
  \bibfield  {author} {\bibinfo {author} {\bibfnamefont {S.}~\bibnamefont
  {Jain}}, \bibinfo {author} {\bibfnamefont {S.}~\bibnamefont {Minwalla}},
  \bibinfo {author} {\bibfnamefont {T.}~\bibnamefont {Sharma}}, \bibinfo
  {author} {\bibfnamefont {T.}~\bibnamefont {Takimi}}, \bibinfo {author}
  {\bibfnamefont {S.~R.}\ \bibnamefont {Wadia}}, \ and\ \bibinfo {author}
  {\bibfnamefont {S.}~\bibnamefont {Yokoyama}},\ }\href {\doibase
  10.1007/JHEP09(2013)009} {\bibfield  {journal} {\bibinfo  {journal} {JHEP}\
  }\textbf {\bibinfo {volume} {09}},\ \bibinfo {pages} {009} (\bibinfo {year}
  {2013})},\ \Eprint {http://arxiv.org/abs/1301.6169} {arXiv:1301.6169
  [hep-th]} \BibitemShut {NoStop}%
\bibitem [{\citenamefont {Chattopadhyay}\ \emph {et~al.}(2018)\citenamefont
  {Chattopadhyay}, \citenamefont {Dutta},\ and\ \citenamefont
  {Dutta}}]{Chattopadhyay:2018wkp}%
  \BibitemOpen
  \bibfield  {author} {\bibinfo {author} {\bibfnamefont {A.}~\bibnamefont
  {Chattopadhyay}}, \bibinfo {author} {\bibfnamefont {P.}~\bibnamefont
  {Dutta}}, \ and\ \bibinfo {author} {\bibfnamefont {S.}~\bibnamefont
  {Dutta}},\ }\href {\doibase 10.1007/JHEP05(2018)117} {\bibfield  {journal}
  {\bibinfo  {journal} {JHEP}\ }\textbf {\bibinfo {volume} {05}},\ \bibinfo
  {pages} {117} (\bibinfo {year} {2018})},\ \Eprint
  {http://arxiv.org/abs/1801.07901} {arXiv:1801.07901 [hep-th]} \BibitemShut
  {NoStop}%
\bibitem [{\citenamefont {Mlawer}\ \emph {et~al.}(1991)\citenamefont {Mlawer},
  \citenamefont {Naculich}, \citenamefont {Riggs},\ and\ \citenamefont
  {Schnitzer}}]{Naculich:1991}%
  \BibitemOpen
  \bibfield  {author} {\bibinfo {author} {\bibfnamefont {E.~J.}\ \bibnamefont
  {Mlawer}}, \bibinfo {author} {\bibfnamefont {S.~G.}\ \bibnamefont
  {Naculich}}, \bibinfo {author} {\bibfnamefont {H.~A.}\ \bibnamefont {Riggs}},
  \ and\ \bibinfo {author} {\bibfnamefont {H.~J.}\ \bibnamefont {Schnitzer}},\
  }\href {\doibase 10.1016/0550-3213(91)90110-J} {\bibfield  {journal}
  {\bibinfo  {journal} {Nucl. Phys.}\ }\textbf {\bibinfo {volume} {B352}},\
  \bibinfo {pages} {863} (\bibinfo {year} {1991})}\BibitemShut {NoStop}%
\bibitem [{\citenamefont {Francesco}\ \emph {et~al.}()\citenamefont
  {Francesco}, \citenamefont {Mathieu},\ and\ \citenamefont
  {S\'en\'echal}}]{yellowbook}%
  \BibitemOpen
  \bibfield  {author} {\bibinfo {author} {\bibfnamefont {P.~D.}\ \bibnamefont
  {Francesco}}, \bibinfo {author} {\bibfnamefont {P.}~\bibnamefont {Mathieu}},
  \ and\ \bibinfo {author} {\bibfnamefont {D.}~\bibnamefont {S\'en\'echal}},\
  }\href@noop {} {\bibinfo  {journal} {Springer}\ }\BibitemShut {NoStop}%
\bibitem [{\citenamefont {Aganagic}\ \emph {et~al.}(2006)\citenamefont
  {Aganagic}, \citenamefont {Neitzke},\ and\ \citenamefont
  {Vafa}}]{Aganagic:2005dh}%
  \BibitemOpen
\bibfield  {journal} {  }\bibfield  {author} {\bibinfo {author} {\bibfnamefont
  {M.}~\bibnamefont {Aganagic}}, \bibinfo {author} {\bibfnamefont
  {A.}~\bibnamefont {Neitzke}}, \ and\ \bibinfo {author} {\bibfnamefont
  {C.}~\bibnamefont {Vafa}},\ }\href {\doibase 10.4310/ATMP.2006.v10.n5.a1}
  {\bibfield  {journal} {\bibinfo  {journal} {Adv. Theor. Math. Phys.}\
  }\textbf {\bibinfo {volume} {10}},\ \bibinfo {pages} {603} (\bibinfo {year}
  {2006})},\ \Eprint {http://arxiv.org/abs/hep-th/0504054}
  {arXiv:hep-th/0504054} \BibitemShut {NoStop}%
\bibitem [{\citenamefont {Okuda}(2005)}]{Okuda}%
  \BibitemOpen
  \bibfield  {author} {\bibinfo {author} {\bibfnamefont {T.}~\bibnamefont
  {Okuda}},\ }\href {\doibase 10.1088/1126-6708/2005/03/047} {\bibfield
  {journal} {\bibinfo  {journal} {JHEP}\ }\textbf {\bibinfo {volume} {03}},\
  \bibinfo {pages} {047} (\bibinfo {year} {2005})},\ \Eprint
  {http://arxiv.org/abs/hep-th/0409270} {arXiv:hep-th/0409270 [hep-th]}
  \BibitemShut {NoStop}%
\bibitem [{\citenamefont {Chattopadhyay}\ \emph {et~al.}(2017)\citenamefont
  {Chattopadhyay}, \citenamefont {Dutta},\ and\ \citenamefont
  {Dutta}}]{Chattopadhyay}%
  \BibitemOpen
  \bibfield  {author} {\bibinfo {author} {\bibfnamefont {A.}~\bibnamefont
  {Chattopadhyay}}, \bibinfo {author} {\bibfnamefont {P.}~\bibnamefont
  {Dutta}}, \ and\ \bibinfo {author} {\bibfnamefont {S.}~\bibnamefont
  {Dutta}},\ }\href {\doibase 10.1007/JHEP11(2017)186} {\bibfield  {journal}
  {\bibinfo  {journal} {JHEP}\ }\textbf {\bibinfo {volume} {11}},\ \bibinfo
  {pages} {186} (\bibinfo {year} {2017})},\ \Eprint
  {http://arxiv.org/abs/1708.03298} {arXiv:1708.03298 [hep-th]} \BibitemShut
  {NoStop}%
\bibitem [{\citenamefont {Migdal}(1983)}]{Migdal:1984gj}%
  \BibitemOpen
  \bibfield  {author} {\bibinfo {author} {\bibfnamefont {A.~A.}\ \bibnamefont
  {Migdal}},\ }\href {\doibase 10.1016/0370-1573(83)90076-5} {\bibfield
  {journal} {\bibinfo  {journal} {Phys. Rept.}\ }\textbf {\bibinfo {volume}
  {102}},\ \bibinfo {pages} {199} (\bibinfo {year} {1983})}\BibitemShut
  {NoStop}%
\bibitem [{\citenamefont {Gopakumar}\ and\ \citenamefont
  {Vafa}(1999)}]{Gopakumar:1998ki}%
  \BibitemOpen
  \bibfield  {author} {\bibinfo {author} {\bibfnamefont {R.}~\bibnamefont
  {Gopakumar}}\ and\ \bibinfo {author} {\bibfnamefont {C.}~\bibnamefont
  {Vafa}},\ }\href@noop {} {\bibfield  {journal} {\bibinfo  {journal} {Adv.
  Theor. Math. Phys.}\ }\textbf {\bibinfo {volume} {3}},\ \bibinfo {pages}
  {1415} (\bibinfo {year} {1999})},\ \Eprint
  {http://arxiv.org/abs/hep-th/9811131} {arXiv:hep-th/9811131 [hep-th]}
  \BibitemShut {NoStop}%
\bibitem [{\citenamefont {Morita}\ and\ \citenamefont
  {Sugiyama}(2017)}]{Morita:2017oev}%
  \BibitemOpen
  \bibfield  {author} {\bibinfo {author} {\bibfnamefont {T.}~\bibnamefont
  {Morita}}\ and\ \bibinfo {author} {\bibfnamefont {K.}~\bibnamefont
  {Sugiyama}},\ }\href@noop {} {\  (\bibinfo {year} {2017})},\ \Eprint
  {http://arxiv.org/abs/1704.08675} {arXiv:1704.08675 [hep-th]} \BibitemShut
  {NoStop}%
\end{thebibliography}%
\end{document}